\documentclass[showpacs,twocolumn,prb]{revtex4-1}
\usepackage{graphicx}
\usepackage{amsmath}
\newcommand{\be}{\begin{equation}}
\newcommand{\ee}{\end{equation}}

\def\beq{\begin{equation}}
\def\eeq{\end{equation}}

\newbox\tablebox    \newdimen\tablewidth
\def\leaderfil{\leaders\hbox to 5pt{\hss.\hss}\hfil}

\def\tablenote#1 #2\par{\begingroup \parindent=0.8em
    \abovedisplayshortskip=0pt\belowdisplayshortskip=0pt
    \noindent
    $$\hss\vbox{\hsize\tablewidth \hangindent=\parindent \hangafter=1 \noindent
    \hbox to \parindent{\sup{\rm #1}\hss}\strut#2\strut\par}\hss$$
    \endgroup}

\begin{document}
\title{The Spectral Decomposition of the Helium atom two-electron configuration in terms of Hydrogenic orbitals}
\author{J. Hutchinson,$^1$ M. Baker$^1$ and F. Marsiglio$^{1,2,\ast}$}
\affiliation{$^1$ Department of Physics, University of Alberta, Edmonton, Alberta, Canada, T6G~2E1
\\
$^2$Physics Division, School of Science and Technology
University of Camerino, I-62032 Camerino (MC), Italy}

\begin{abstract}

The two electron configuration in the Helium atom is known to very high precision.
Yet, we tend to refer to this configuration as a $1s\uparrow 1s\downarrow$ singlet, where
the designations refer to Hydrogen orbitals. The high precision calculations utilize
basis sets that are suited for high accuracy and ease of calculation, but do not really
aid in our understanding of the electron configuration {\em in terms of product
states of Hydrogen orbitals}. Since undergraduate students are generally taught to
think of Helium, and indeed, the rest of the periodic table, in terms of Hydrogenic 
orbitals, we present in this paper a detailed spectral decomposition of the two 
electron ground state for Helium in terms of these basis states. The 
$1s\uparrow 1s\downarrow$ singlet contributes less than $93\%$ to the ground state
configuration, with other contributions coming from both bound and continuum Hydrogenic states.

\end{abstract}

\date{\today} 
\maketitle

\section{Introduction}

As early as 1928 Hylleraas\cite{hylleraas29a,hettema00}  recognized that using Hydrogenic orbitals to describe the ground state
electron configuration of the Helium atom was not a good idea. In the introduction to his seminal paper
he says, ``It thereby appeared that the use of hydrogen eigenfunctions, as done by Dr. Biem\"uller, leads to erroneous results, which obviously has to do with the fact that they do not form a complete functional system.''\cite{hylleraas29a} 
He then goes on to introduce a set of coordinates and variational wave functions that more accurately describe the 
two-electron ground state of the Helium atom.\cite{hylleraas29a,hylleraas29b}

To our knowledge, the results of Dr. Biem\"uller were never published. Most students nowadays, if asked about the
electronic ground state configuration of the Helium atom, will respond that 
it is the singlet state\cite{rem1} of $1s\uparrow 1s\downarrow$,
as listed, for example, in many periodic tables. Even more senior colleagues, while realizing that this simple characterization 
omits interactions, become disturbed at the suggestion that other orbitals enter the ground state configuration, since this 
appears, at first glance, to be at odds with the `closed-shell' character of He, and the fact that it is an `inert' element.
This misunderstanding tends to be reinforced in the way we teach undergraduate quantum mechanics: since the 
two-electron problem is already very difficult, Helium is often used as an example case study for many approximate 
methods, including perturbation theory and the variational method. Usually only corrections to the energy 
(and not the wave function) are described, and even with the variational method, often a simple extension of a 
(singlet) $1s\uparrow 1s\downarrow$ is used. This tends to {\em reinforce} the
incorrect idea that the singlet $1s\uparrow 1s\downarrow$ Hydrogenic product wave function 
is the entire picture.\cite{rem2}

The solution to this problem attained by Hylleraas was to adopt a set of basis functions that more accurately 
capture the electron correlations that exist in the Helium two-electron ground state. We want to emphasize here 
that there is no question that this works {\em extremely well}.\cite{schwartz06} Nonetheless, physicists and chemists 
still tend to think of atomic electronic configurations in terms of Hydrogenic orbitals (the periodic table guides us in 
this direction), and this is how our intuition is formed. Thus, in spite of the lesson of Dr. Biem\"uller's (erroneous) calculation, 
we think it is important to answer the question, `Just how much of the Helium two-electron ground state consists of the 
singlet $1s\uparrow 1s\downarrow$ configuration?' and, as an obvious follow-up, `What other Hydrogenic states contribute 
to the ground state configuration?'\cite{rem3} Most (non-expert) colleagues are shocked when the answer requires
continuum states.\cite{rem4}

Aside from pedagogical value, the requirement of more than one electronic configuration has also become a 
potentially important ingredient in models of metals and superconductors within the field of condensed 
matter.\cite{hirsch01,bach10,bach12} The electronic ground state of Helium has served as the `poster child' for 
atoms in a solid state environment, where the electron occupation can fluctuate. An important ingredient in 
electron conduction is the fact that most lattice models in condensed matter physics focus on one set of orbitals 
from each atomic site; these orbitals overlap to form a conduction band, and the band that crosses the Fermi energy,
i.e. the energy separating occupied from unoccupied bands, is the one band of interest for low energy 
excitations. But this reasoning is primarily based on single electron ideas, and as a remedy, this model 
(tight-binding, or H\"uckel) is often generalized to include electron interactions with one another, especially when 
two electrons occupy the same orbital on the same site (the so-called Hubbard model\cite{hubbard63}). 
The Hubbard model,  and, in general, Hubbard-like models, are the subject of intensive investigation in 
condensed matter.\cite{rem5} However, almost all these models miss the important ingredient that the same 
single electron orbitals that are pertinent for single electron occupation are {\em not adequate} to describe a 
doubly occupied situation. That is, the situation, ``when two electrons occupy the same orbital on the same 
atom'' is simply not possible, without a modification of those orbitals.\cite{rem6} In reality, two electrons that find themselves 
on the same atom will spread out to avoid one another, and therefore inevitably occupy (at least partially) other single
particle orbitals (like the $2s$ and $3s$ orbitals in the case of Helium). The present study of Helium, while avoiding more complicated ideas like Wannier bases and Hartree-Fock orbitals, will emphasize this 
very point. As such, the two-electron configuration in the Helium atom serves as a stepping stone between a 
standard quantum mechanical `textbook' problem and a more research-oriented set of problems.

We first outline the general problem of solving for the eigenstates of two-electrons bound to a central nucleus. This will first
be done using `natural' basis states consisting of product states of the bound state hydrogenic states. This is presumably the path
followed by Dr. Biem\"uller, but since, to our knowledge, this exercise has never been published, we provide an outline
of the method, with most of the details relegated to an appendix. As already remarked, this procedure does {\em not} succeed in
converging to the known answers for the Helium atom, for the reason that the continuum states are required, and we
arrive at the surprising (for some) conclusion that an accurate description of the Helium ground state {\em cannot} be
provided by using just the Hydrogenic bound states.

We thus turn to a more direct approach, where, starting with the exact wave function, one can compute the coefficients of the
various basis states of interest. 
For Helium, while the exact ground state wave function is known numerically to many significant 
digits,\cite{schwartz06} we instead use the so-called Hylleraas wave function,\cite{hylleraas29b} consisting of only three variational parameters,\cite{rem7} in the interest of simplicity and transparency. This will serve as our `exact' wave
function, and, consistent with what we stated above, the sum of the magnitudes of the coefficients of the basis states
consisting of product Hydrogenic orbitals will fall short of unity, with the remainder coming from the continuum
states.

While much of this discussion is suited for undergraduates, some of the mathematics is more suited for 
undergraduate projects and/or graduate courses. In either case we feel that this approach to the problem provides
a useful connection between course work and research-type problems. We have tried to keep much of the mathematics
(which we have done in part analytically and in part with software packages like Mathematica or Maple) in Appendices.

\section{the Helium two electron problem}

\subsection{Preliminaries}

The energetics of the two electron Helium atom are very well known.\cite{schwartz06} 
Here we concern ourselves with just the non-relativistic interactions, so
the Hamiltonian governing this system is
\beq
H=\frac{-\hbar^2}{2m}(\nabla_1^2+\nabla_2^2)-\frac{Ze^2}{4\pi\epsilon_0}\bigg(\frac{1}{r_1}+\frac{1}{r_2}\bigg)+\frac{e^2}{4\pi\epsilon_0}\frac{1}{r_{12}}
\label{ham}
\eeq
where $r_{12} \equiv | \vec{r_1} - \vec{r_2}|$ is the separation between the two electrons, and $r_1 \equiv |\vec{r_1}|$ ($r_2 \equiv |\vec{r_2}|$). Note that
we have already used an important approximation in writing down Eq. (\ref{ham}) --- we have adopted the Born-Oppenheimer approximation, which
essentially assumes that the mass of the nucleus is infinite. Hence the only degrees of freedom are those of the two electrons, each with charge $-e$ and
mass $m$. The factor of $Z$ in the middle term is present because in general the charge of the nucleus is $Ze$;
obviously for Helium $Z=2$.
It is the last term, representing the electron-electron repulsion, that causes the difficulty; this is the term which is often ignored (at least conceptually) in
a student's first exposure to the Helium atom (and the periodic table, for that matter) in his/her undergraduate education.

The usual procedure is to first ignore the electron-electron repulsion; the problem is then readily solved since it now consists of essentially a Hydrogen 
problem that has to accommodate two electrons. The ground state solution is then
\begin{equation}
\psi_1(\vec{r_1},\vec{r_2}) = \phi_{100}(\vec{r_1}) \phi_{100}(\vec{r_2}),
\label{soln0}
\end{equation}
which is a product state of two single electron solutions,\cite{levine09,griffiths05}
\begin{equation}
\phi_{\rm n\ell m}(\vec{r}) = R_{\rm n\ell}(r) Y_{\ell}^m(\theta,\phi),
\label{soln_nlm}
\end{equation}
where $(n\ell m)$ are the usual quantum numbers, and $R(r)$ [$Y_{\ell}^m(\theta,\phi)$] is the radial [angular] part of the wave function, which can
be written in terms of Associated Laguerre polynomials and Spherical Harmonics, respectively.
These functions are tabulated in most texts; for reference we write down the ground state result needed in Eq. (\ref{soln0}):
\begin{equation}
\phi_{100}(\vec{r_1}) = 2 \bigl({Z \over a_0}\bigr)^{3/2} e^{-Zr_1/a_0} {1 \over \sqrt{4 \pi}}
\label{soln_100}
\end{equation}
where $a_0$ is the Bohr radius and it is understood that $Z = 2$ for Helium, but we have left it general in Eq. (\ref{soln_100}).

As it stands, Eq. (\ref{soln0}) is deceptively simple; we have left out the spin degree of freedom, and therefore we 
have omitted any discussion of {\em parahelium} states (symmetric spatial and antisymmetric spin wave 
function components) and {\em orthohelium} (antisymmetric spatial and symmetric spin wave
function components). In the absence of spin-orbit coupling these two classes of states are {\em not} 
coupled by the Hamiltonian (\ref{ham}), and one can focus on one or the other. Since we are interested 
in the ground state configuration, we will focus on the parahelium states --- these contain the lowest energy basis state and
therefore the ground state. This means
that all the two particle states that we consider below should be understood to include the singlet spin state,
\begin{equation}
|\chi(\vec{s}_1,\vec{s}_2) \rangle = {1 \over \sqrt{2}} \bigl[|\uparrow\rangle_1 \ | \downarrow \rangle_2 - |\downarrow\rangle_1 \ | \uparrow \rangle_2 \big],
\label{singlet}
\end{equation}
where the subscripts $1$ and $2$ refer to particle $1$ and $2$.\cite{rem8} From this point on spin is no longer 
included in the discussion, but its presence has indeed dictated that we treat only {\em symmetric} orbital states, 
of which the state (\ref{soln0}) is one. Other examples are listed in the Appendix, and will be used below.

\subsection{Matrix Mechanics}

Because this approach was so successful in other problems, our first line of attack is to utilize matrix 
mechanics; we decompose the ground state into a complete set of simple well known basis states,
\beq
|\psi\rangle=\sum^\infty_{i=1}a_i|\psi_i\rangle
\label{decompose}
\eeq
where the $a_i$'s are the unknown coefficients. Inserting this into the time-independent Schr\"odinger equation, and taking inner products with the
same basis states yields the eigenvalue equation,
\be
\sum_{j=1}^\infty H_{ij} a_j = E a_i,
\label{eig_eqn}
\ee
where the matrix elements are given by
\begin{equation}
H_{ij} = \langle \psi_i|H|\psi_j \rangle.
\label{HM}   
\end{equation}
As already remarked, a {\em complete} basis set of Hydrogenic orbitals consists of an infinite number of bound 
states (of which Eqs. (\ref{e1}-\ref{e6}) are the first few), {\em and} an infinite continuum of unbound states, for 
which a much more complicated expression is required (though it is essentially the analytical continuation of
the bound states). But our tact will be to forge ahead, and simply truncate the expansion to include only the 
low-lying bound states, thus excluding a (infinite) number of bound states and all the continuum states. Such a 
truncation scheme worked extremely well for the harmonic oscillator,\cite{marsiglio09}
where, in the square well basis used for that problem, only the first ten or so 
states were required to give fully converged results.

One calculational advantage of this choice of basis states is that the one electron parts 
of the Hamiltonian [all but the last term in Eq. (\ref{ham})] return eigenvalues of
the Hydrogen spectrum when operating on these states. That is, focusing on a 
product, $\phi_{n_1,\ell_1,m_1}(\vec{r_1})  \phi_{n_2,\ell_2,m_2}(\vec{r_2})$,
which covers the most general case encountered in Eqs. (\ref{e1}-\ref{e6}), we have
\begin{eqnarray}
&&H \phi_{n_1,\ell_1,m_1}(\vec{r_1})  \phi_{n_2,\ell_2,m_2}(\vec{r_2}) = \nonumber \\
&&\bigl[ -Z^2E_0 \bigl({1 \over n_1^2} + {1 \over n_2^2}\bigr) + \hat{H}_{\rm int} \bigr]
 \phi_{n_1,\ell_1,m_1}(\vec{r_1})  \phi_{n_2,\ell_2,m_2}(\vec{r_2}),\nonumber \\
 &&
\label{hop}
\end{eqnarray}
where
\be
\hat{H}_{\rm int} = {e^2 \over 4 \pi \epsilon_0}{1 \over |\vec{r_1} - \vec{r_2}|}
\label{hint}
\ee
is the ``hard part'', i.e. the electron-electron interaction potential.
In Eq. (\ref{hop}) and below we use $Z = 2$ and $E_0 \equiv \hbar^2/(2ma_0^2) \approx 13.606$ eV.
To consider an actual matrix element between any general two electron states, we require twelve quantum numbers; hence to avoid proliferation
of indices we use the shorthand $i_1 \equiv n_1 \ell_1 m_1$. Then
\begin{eqnarray}
\langle \phi_{i_1} \phi_{i_2} | H | \phi_{i_3} \phi_{i_4} \rangle &=& -Z^2 E_0\delta_{i_1,i_3} \delta_{i_2,i_4}  
\bigl( {1 \over n_1^2} + {1 \over n_2^2} \bigr) \nonumber \\
&+& \langle \phi_{i_1} \phi_{i_2} | H_{\text int} | \phi_{i_3} \phi_{i_4} \rangle,
\label{mat_ele}
\end{eqnarray}
where the last matrix element requires more careful analysis; details of the calculation of this matrix element, including
some examples, are provided in Appendix A. There a number of states are listed [see Eqs. (\ref{e1}-\ref{e6})], and
a software package like Maple or Mathematica can be tasked with evaluating all the required matrix elements 
up to some cutoff, $i_{\rm max}$; the resulting finite matrix can be easily diagonalized, and we can obtain the corresponding 
ground state eigenvalue and eigenvector.

\subsection{Results} 

\begin{figure}[h!]
\begin{center}
\includegraphics[height=3.1in,width=2.6in, angle=-90]{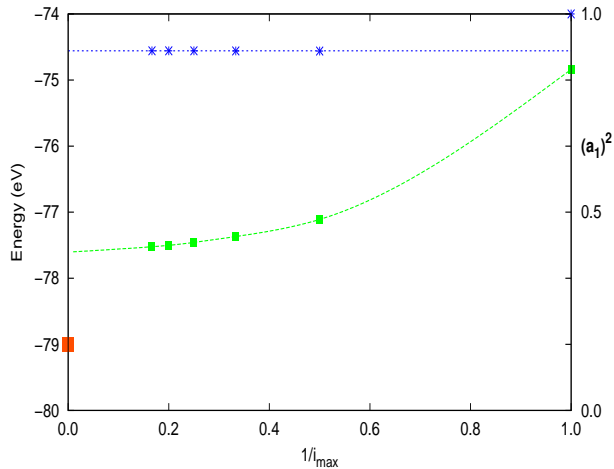} 
\caption{(color online)
Results for the ground state energy (filled green squares) as a function of 
$i_{\rm max}$ (see text and Appendix A for a full explanation of $i_{\rm max}$). The 
curve provided is merely a guide to the eye. Notice, however, that the extrapolated 
result to the origin (as $i_{\rm max} \rightarrow \infty$) is well above the known
exact result, indicated with the large (red) square at $-79.0$ eV. The discrepancy 
is explained in the text. Also shown are blue asterisks, which indicate
the probability of the $(1s \uparrow 1s \downarrow)$ basis state (right hand scale). It 
saturates at approximately $(a_1)^2 \equiv a^2_{(100,100)} \approx 0.91$, through
which a horizontal (blue) line is drawn, indicating that
a significant amplitude comes from other states.}
\label{fig1}
\end{center}
\end{figure}

In Fig.~1 we show the ground state energy as a function of the inverse of 
$i_{\rm max}$, with square symbols, along with a guide to the eye; it is clear that
the energy has essentially saturated as $i_{\rm max}$ increases, but to a value 
much higher than the known exact value, which is indicated by the large (red)
square at the origin. The reason for this discrepancy has already been noted: 
including only the bound states is not sufficient --- they do not form a complete
basis set, and the contribution from the continuum states is significant. At first glance it 
may seem odd that continuum states contribute to the ground state for the
Helium atom; however, one must keep in mind that (i) the right combination 
of plane waves, for example, can indeed describe a very localized state, and 
(ii) with a poor choice of basis states (which, indeed, the Hydrogenic states 
are), the bound states will simply not be sufficient to describe the Helium atom. This 
latter point is important, as a truncation in the (infinite) basis for the problem in 
Ref. (\onlinecite{marsiglio09}) did not deteriorate the solution there, as the set of basis 
states used in that reference were sufficiently `good' to describe the problem at hand. 
Later we shall demonstrate the role of continuum states in this problem.

Also shown in Fig.~1 is the probability for the singlet $(1s \uparrow 1s \downarrow)$ basis 
state (right hand scale) as  a function of $1/i_{\rm max}$ (blue asterisks). 
The horizontal line indicates the saturated value, about $91\%$, indicating that 
a significant fraction (close to $10\%$) comes from other contributions. However,
because of the problem mentioned above, this value is not reliable, and we will 
tackle this issue in the next section. 

To summarize this section, we have attempted to describe the configuration
of the two-electron ground state in Helium, using only the bound state Hydrogenic
orbitals. The reason for doing this is that these orbitals are the ones with which we
have the most intuition, and they are the ones we tend to use to gain a preliminary understanding
of the periodic table. Our hope was that, while this set of basis states is infinite, a 
finite set used through some truncation scheme would capture the essence of the 
ground state configuration; in reality not only did this not work well, it is clear that
even if one extrapolates to the use of all the Hydrogenic bound states, this incomplete
basis will not work.

In fact, the continuum states are required for a proper description. The reason for this
is that the continuum states are able to describe high resolution spatial correlations that
the bound states cannot. The bound states become more extended as their quantum numbers
increase, and so any finer scale spatial correlations will have to be properly described by
the continuum states. This conclusion will
be reinforced in the next section, where we examine projections of the various basis
states on a very accurate but simple representation of the exact two-electron wave function.

\section{Projections}

\subsection{Hylleraas wave function: bound states}

One could try to include the continuum states in the preceding calculation. To avoid 
an infinite matrix would require a judicious selection of these states, presumably based 
on their energies. In our opinion this procedure is fraught with difficulties and various
choices are possible, so we do not pursue this tact. What we really want is a good 
estimate of the contribution of the `naive' $(1s \uparrow 1s \downarrow)$ basis state 
to the actual electronic ground state of Helium. The contributions from the
continuum states are of importance only insofar as they influence the value 
$a_1^2 \approx 0.91$ obtained in the previous section. Thus we take
a different approach, and find that the conclusion of the previous section is 
reinforced quantitatively.
 
Using Eq. (\ref{decompose}), we can formally rewrite this as
\beq
\Psi_{\rm exact}=\sum^{\infty}_{m=0}a_m|\psi_m\rangle + \int \ dp \ a_p|\psi_p^{\rm cont}\rangle,
\label{decomp}
\eeq
where now it is clear that a complete set of states is being used; the $|\psi_m\rangle$ 
are those enumerated in Eqs. (\ref{e1}-\ref{e6}) and beyond, while the $|\psi_p^{\rm cont}\rangle$ 
refer to the (as yet unspecified) continuum basis states. These states, taken together, in fact form a 
complete, orthonormal set. Multiplying on the left by $\langle\psi_n|$ and forming 
the inner product therefore projects out the contribution from the $n^{\rm th}$ basis state:
\beq
a_n = \int\psi_n \Psi_{\rm exact}.
\label{a_n}
\eeq

For Eq. (\ref{a_n}) to be useful we need to know $\Psi_{\rm exact}$; as mentioned in 
the Introduction, this is known numerically to many digits precision,\cite{schwartz06,rem7} 
using a Hylleraas-type basis:\cite{hylleraas29b,schmid87}
\be
\Psi = \sum c_{i,j,\ell} e^{-ks/2} s^j \ u^\ell \ t^i
\label{hylla}
\ee
where the Hylleraas coordinates are defined 
\be
s \equiv (r_1 + r_2)/a_0, \ \ \ t \equiv (r_2 - r_1)/a_0, \  \ \ u \equiv r_{12} \equiv |\vec{r_1} - \vec{r_2}|/a_0
\label{coord}
\ee
and the summation occurs all over non-negative integers for $j$ and $\ell$, and 
only even non-negative integers for $i$. This variational basis set has been expanded 
in a variety of ways, as summarized in Ref. (\onlinecite{schwartz06}), but in what 
follows we take only three terms from Eq. (\ref{hylla}), with 
$(i,j,\ell) = (0,0,0)$, $(i,j,\ell) = (2,0,0)$, and $(i,j,\ell) = (0,0,1)$. 
Then we define the `Hylleraas wave function' to be the three parameter wave function,
\be
\Psi_{\rm Hy}(\vec{r_1},\vec{r_2}) = {2 \over \pi} \biggl({kZ \over a_0}\biggr)^3 e^{-Zks}\bigl[1+2Zc_1ku + c_2(2Zkt)^2 \bigr],
\label{hy}
\ee
where $k$, $c_1$, and $c_2$ are the three parameters to be determined 
by minimizing the energy, and $Z$ will eventually be set to equal 2. Eq. (\ref{hy})
will serve as our `exact' wave function. In reality, minimization of the energy 
with $\Psi_{\rm Hy}$ yields $E_{\rm Hy} \approx -78.979$ eV, with variational
parameters $c_1\approx 0.0803$ and $c_2 \approx 0.0099$, and $k \approx 0.908$. 
Evaluating the necessary integrals and obtaining these results is
well documented,\cite{hylleraas29b,schmid87}  and so these are simply summarized in Appendix B.
The attained energy is within $0.05\%$ of the exact result ($-79.014$ eV); we
consider this sufficiently close to justify our adoption for the present purposes 
of ${1 \over \sqrt{N}}|\Psi_{\rm Hy}\rangle$ as the `exact' normalized Helium 
wave function, with $N$ given below. The calculation of the overlap integrals for the bound states
can be done for the general case --- some details are provided in Appendix B, along with a simple
example. Our results are summarized in Table 1.

\begin{table}
\caption{Results for some overlaps, $a_i$.}
\begin{tabular}{|l|l|l|l|l|}
\hline
$i$ & basis state & $a_i$ & $|a_i|^2$ & Tot. Prob.\\
\hline
1 & 100 100 &   0.9624 & 0.9263 & 0.9263 \\
2 & 100 200 & -0.2148 & 0.0461 & 0.9725 \\
3 & 100 300 & -0.0752 & 0.0057 & 0.9781 \\
4 & 100 400 & -0.0427 & 0.0018 & 0.9799 \\
5 & 100 500 & -0.0289 & 0.0008 & 0.9807 \\
6 & 100 600 & -0.0213 & 0.0005 & 0.9812 \\
7 & 100 700 & -0.0166 & 0.0003 & 0.9815 \\
8 & 21-1 211 & 0.0260 & 0.0007 & 0.9822 \\
9 & 210 210 & -0.0184 & 0.0003 & 0.9825 \\
10 & 200 200 & -0.0146 & 0.0002 & 0.9827 \\
11 & 200 300 & -0.0090 & 0.0001 & 0.9828 \\
12 & 100 320 &  0           & 0          & 0.9828 \\
\hline 
\end{tabular}
\end{table}

It is clear that the largest contributions arise from states in which one of the electrons is 
in the $\phi_{100}(\vec{r})$ state, i.e. the single electron ground state. In particular, more than $92\%$ comes
from the $1s\uparrow 1s\downarrow$ singlet, but a sizeable contribution comes from states other than this one.
Further (small) contributions arise from states not listed. While the total 
probability (5th column in Table~1) is close to unity, inclusion of the
remaining bound states (not shown) still gives a total probability that falls short of $0.99$. The
remaining probability arises from continuum states, to which we turn in the next subsection.

\subsection{Overlap integrals: continuum states}

For Helium, including the full spectrum of continuum two particle states into the problem of computing
the ground state configuration is very complicated. Noting that the most important contributions
from the bound states arise when one of the electrons is in the one electron ground state (see Table~1),
we will likewise compute only the contributions from the continuum states when one of the
two electrons is in the one electron (bound) ground state. These states, 
in the singlet configuration, are written
\begin{equation}
\psi_p(\vec{r_1},\vec{r_2}) = {1 \over \sqrt{2}}\bigl( \phi_p(r_1) \phi_{100}(r_2) + 
\phi_{100}(r_1) \phi_p(r_2) \bigr)
\label{scatt}
\end{equation}
where only one label (the momentum $p \equiv |\vec{p}|$) is required because the other electron is in the
$1s$ state so the continuum state also has $\ell = m = 0$, and we have written the
right-hand-side of Eq. (\ref{scatt}) as depending only on the radial coordinates, $r_1$ and
$r_2$. The single particle state $\phi_p(r)$ has a radial part  $R_p(r)$ given by
\begin{equation}
R_p(r) = {Z\over a_0} \sqrt{2 \pi {p a_0 \over Z} \over 1 - e^{-2\pi {Z \over pa_0}}} 
e^{-ipr} M(1 + i{Z \over pa_0}, 2, 2ipr),
\label{cont_radial}
\end{equation}
where we have written this for general $Z$ (here we need only $Z=2$), and
\begin{equation}
M(a,b,z) \equiv \sum_{m=0}^\infty {(a)_m \over (b)_m}{z^m \over m!}
\label{kummer}
\end{equation}
is the so-called Kummer function,\cite{abramowitz72} and 
$(a)_m \equiv a(a+1)(a+2)...(a+m-1)$ is the so-called Pochhammer
symbol. Note that $(a)_0 \equiv 1$. Equation (\ref{cont_radial}) is 
essentially the analytical continuation of the radial bound state wave 
functions. 
The standard\cite{landau65} normalization condition for the 
continuum states,
\begin{equation}
\int_0^\infty \ dr \ r^2 R_{p^\prime}(r) R_{p}(r) = \delta (p^\prime - p),
\label{norm}
\end{equation}
determines the coefficient in Eq. (\ref{cont_radial}). Just as for the bound states we require
\begin{equation}
a_p = \int d^3r_1 \int d^3r_2 \psi_p(r_1,r_2) {1 \over \sqrt{N}} \Psi_{Hy}(\vec{r_1},\vec{r_2}).
\label{expansion}
\end{equation}
Because of symmetry the contributions from the two terms in Eq. (\ref{scatt}) are identical;
the integrals involved in Eq. (\ref{expansion}) can be done, either with Mathematica or by hand.
Some detail is provided in Appendix~C. With $y \equiv pa_0/Z$ the result is
\begin{widetext}
\begin{eqnarray}
a_p  =  \sqrt{\pi \over N}{32k^3 \over (k+1)^5}\sqrt{a_0 \over Z} \sqrt{y \over 
1 - e^{-2 \pi \over y}} \biggl\{
&&2(k+1)^2 I_2+ \nonumber \\
&c_1& \bigl[ 16kI_1 + 4k(k+1)^2I_3 -4k(k+1)J_2 - 16kJ_1 \bigr] + \nonumber \\
&c_2& \bigl[ 96k^2I_2 + 8k^2(k+1)^2I_4 - 48k^2(k+1)I_3 \bigr] \biggr\}
\label{expansion2}
\end{eqnarray}
\end{widetext}
where the expression for $I_n$ and $J_n$ are provided in Appendix C.
Then the total contribution for these continuum states,
\begin{equation}
P_{\rm cont}  = {2 \over \pi}\int_0^\infty dp |a_p|^2,
\label{total_cont}
\end{equation}
is evaluated numerically; the result is a further contribution of $P_{\rm cont} \approx 0.0117$,
which brings the total probability from the bound states in Table 1 and the continuum
states shown in Eq. (\ref{scatt}) to $\approx 0.995$. The remaining probability required to reach unity comes
from bound state contributions not included in Table 1 and from continuum states beyond
those not considered in Eq. (\ref{scatt}). 

\section{Summary}

The primary purpose of this paper was to demonstrate the degree to which the electron
configuration in the Helium atom is {\em not} simply the singlet $1s\uparrow 1s\downarrow$ state.
In fact we have shown, through two methods, diagonalization in a particular basis set,
and through projection on this same basis, that approximately 8\% of the wave function is
not the $1s\uparrow 1s\downarrow$ singlet. As already noted, this has a very large effect on the energy.
While we mentioned in the introduction that this
fact may be of interest in current research in strongly correlated electron systems, including
superconductors, the emphasis here has been on pedagogical aspects. Further
exploration of the consequences of this electron configuration in a solid can be
the topic of specialized student research projects.

The diagonalization of the problem in a truncated basis was, in fact, not
successful at reproducing the ground state energy; in being unsuccessful, this calculation has
served to highlight important pedagogical points, and it is
partly for this reason that we have included a detailed analysis of the problem here. The
two-electron problem in Helium served to highlight that the bound Hydrogenic states
do not form a complete basis set, and while one might have thought that the low-lying
states would be sufficient to accurately describe the ground state, this work shows
that this supposition is incorrect. 

Setting up the electron configuration in Helium as a 
matrix diagonalization problem also serves to provide a concrete example
of matrix mechanics, which, for undergraduates at least, is often introduced only in a
formal sense, with an abstract-only exposition of Hilbert space and expansion in basis
functions, etc. This problem also serves to provide a first exposure to a realistic `many-body'
problem, and how one would construct many particle wave functions (here, two electrons
is the first stepping stone in this direction beyond the single particle problem to which
undergraduates are normally exposed).

The second method we presented, projection of an accurate wave function 
onto an orthonormal basis set of `known' and well-understood wave functions, requires
a little more sophistication, because Hylleraas wave functions are usually only
introduced in graduate studies. Nonetheless, the three-parameter Hylleraas wave
function given in Eq. (\ref{hy}) is sufficiently simple to be suitable for undergraduates.
The ensuing algebra for the bound states is also readily accessible, particularly if
projections are computed on a case-by-case basis. For example, evaluation of Eq. (\ref{overlap2}) 
requires knowledge only of elementary functions and standard integrations. We have included
projections involving the continuum states as well, and these require more advanced
mathematics, and certainly can be skipped at the undergraduate level. 

The topic of even two electron correlations has been traditionally confined to more
advanced studies at the graduate level. Part of the reason for this is that there are
better and more accurate methods available after a preliminary introduction to
many-body methods has been assimilated by the student. However, most students see the
electronic structure of atoms in the periodic table in general, and the Helium atom in particular, 
in terms of Hydrogen orbitals, and therefore it is desirable that a description be provided in 
terms of these same orbitals, as we have presented in this paper.

\begin{acknowledgments}
We thank Don Page and Bob Teshima for helpful discussions concerning the treatment of the continuum states .This work was supported in part by the Natural Sciences and Engineering Research Council of Canada (NSERC), by the Canadian Institute for Advanced Research (CIfAR), and by a University of Alberta McCalla Professorship and Teaching and Learning Enhancement Fund (TLEF) grant. 

\end{acknowledgments}
\medskip
\noindent $^\ast$permanent address for FM is University of Alberta.
\medskip
\begin{widetext}
\vskip6.0in
\end{widetext}


\appendix
\section{Matrix mechanics for Helium}

\subsection{General considerations}

The first 6 two particle states that need to be considered [dropping the spin part given in Eq. (\ref{singlet})] are, for example:
\begin{widetext}
\begin{subequations}
\begin{align}
\psi_1(\vec{r_1},\vec{r_2}) &=& \phi_{100}(\vec{r_1}) \phi_{100}(\vec{r_2}), && m_1+m_2 = 0
\label{e1}
\\
\psi_2(\vec{r_1},\vec{r_2}) &=& \frac{1}{\sqrt{2}} \bigl[\phi_{100}(\vec{r_1}) \phi_{200}(\vec{r_2}) +
\phi_{100}(\vec{r_2}) \phi_{200}(\vec{r_1}) \bigr],  & & m_1+m_2 = 0
\label{e2}
\\
\psi_3(\vec{r_1},\vec{r_2}) &=& \frac{1}{\sqrt{2}} \bigl[\phi_{100}(\vec{r_1}) \phi_{21-1}(\vec{r_2}) +
\phi_{100}(\vec{r_2}) \phi_{21-1}(\vec{r_1}) \bigr], && m_1+m_2 = -1
\label{e3}
\\
\psi_4(\vec{r_1},\vec{r_2}) &=& \frac{1}{\sqrt{2}} \bigl[\phi_{100}(\vec{r_1}) \phi_{210}(\vec{r_2}) +
\phi_{100}(\vec{r_2}) \phi_{210}(\vec{r_1}) \bigr], && m_1+m_2 = 0
\label{e4}
\\
\psi_5(\vec{r_1},\vec{r_2}) &=& \frac{1}{\sqrt{2}} \bigl[\phi_{100}(\vec{r_1}) \phi_{211}(\vec{r_2}) +
\phi_{100}(\vec{r_2}) \phi_{211}(\vec{r_1}) \bigr], && m_1+m_2 = 1
\label{e5}
\\
\psi_6(\vec{r_1},\vec{r_2}) &=& \phi_{200}(\vec{r_1}) \phi_{200}(\vec{r_2}), && m_1+m_2 = 0
\label{e6}
\end{align}
\end{subequations}
\end{widetext}
Note that all these states are symmetric, and that we have indicated the quantum number $m_1+m_2$ for each 
two-particle state. In any state in which this is not zero, that state will {\em not} contribute to the ground state. 
This selection rule will be derived below, along with some other rules that eliminate more of these states. 
These rules provide a big simplification, and to the degree that we anticipate that maybe the $(n\ell m) = (600)$ 
doesn't contribute very much to the ground state, the number of states that may be required might be anticipated
to be small. Unfortunately, this is not the case, as the continuum states contribute as well; this fact ultimately 
thwarts this attempt to attain a spectral decomposition in terms of these basis states.\cite{hylleraas29b} 
Nonetheless, we proceed along this line of investigation, first to demonstrate that this is the case, 
and secondly to establish these helpful selection rules. 

To evaluate the last matrix element in Eq. (\ref{mat_ele}) we introduce the expansion\cite{jackson75,slater60}
\begin{widetext}
\be
{1 \over |\vec{r_1}-\vec{r_2}|} = \sum_{k=0}^\infty \sum_{m_k=-k}^k  {(k-|m_k|)! \over (k+|m_k|)!} {r^k_< \over r^{k+1}_>} P_k^{|m_k|}(cos\theta_1)P_k^{|m_k|}(cos\theta_2) e^{im_k (\phi_1-\phi_2)}
\label{multipole}
\ee
\end{widetext}
where $r_< = r_1$ and $r_> = r_2$ if $r_1 < r_2$ and vice versa for $r_1 > r_2$.
The $P_k^{m_k}$ are the associated Legendre polynomials, and the angles correspond to the spherical coordinates for each of the vectors, $\vec{r_1}$ and
$\vec{r_2}$.  
Eq. (\ref{multipole}) should become familiar to students through problems in Electromagnetism as well as in
Quantum Mechanics.
We can expand each of the individual Hydrogenic states (see Eq. (\ref{soln_nlm})) and use 
\be
Y_\ell^m(\theta,\phi) = \epsilon_m\sqrt{{(2\ell + 1) \over 4 \pi}{(\ell - |m|)! \over (\ell + |m|)!}} e^{im\phi} P_\ell^m(\cos{\theta}),
\label{aleg}
\ee
where $\epsilon_m = (-1)^m$ for $m \ge 0$, and $\epsilon_m = 1$ for $m \le 0$. If we 
substitute into the last term of Eq. (\ref{mat_ele})) we obtain
\begin{widetext}
\begin{eqnarray}
& &\langle \phi_{i_1} \phi_{i_2} | H_{\text int} | \phi_{i_3} \phi_{i_4} \rangle =  \sum_{k=0}^\infty \sum_{m_k=-k}^k  {(k-|m_k|)! \over (k+|m_k|)!} (-1)^{(m_1+|m_1|+m_2+|m_2|+m_3+|m_3|+m_4+|m_4|)/2}
\nonumber \\
& & \times \sqrt{(2l_1+1)(l_1-|m_1|)! \over (l_1+|m_1|)!} \sqrt{(2l_2+1)(l_2-|m_2|)! \over (l_2+|m_2|)!} \sqrt{(2l_3+1)(l_3-|m_3|)! \over (l_3+|m_3|)!} \sqrt{(2l_4+1)(l_4-|m_4|)! \over (l_4+|m_4|)!}
\nonumber \\
& & \times \int_0^\infty \int_0^\infty R_{n_1,l_1}(r_1) R_{n_2,l_2}(r_2) R_{n_3,l_3}(r_1) R_{n_4,l_4} (r_2) {r^k_< \over r^{k+1}_>} r_1^2 r_2^2 dr_1 dr_2
\nonumber \\
& & \times {1 \over 2} \int_0^\pi P_{l_1}^{|m_1|}(cos\theta_1) P_{l_3}^{|m_3|}(cos\theta_1) P_k^{|m_k|}(cos\theta_1) sin\theta_1 d\theta_1
\nonumber \\
& & \times {1 \over 2} \int_0^\pi P_{l_2}^{|m_2|}(cos\theta_2) P_{l_4}^{|m_4|}(cos\theta_2) P_k^{|m_k|}(cos\theta_2) sin\theta_2 d\theta_2
\nonumber \\
& & \times {1 \over 2\pi} \int_0^{2\pi} e^{i(-m_1+m_3+m_k)\phi_1}d\phi_1 \times {1 \over 2\pi} \int_0^{2\pi} e^{i(-m_2+m_4-m_k)\phi_2}d\phi_2.
\label{test}
\end{eqnarray}
\end{widetext}

\subsection{Simplifications and Selection Rules}

The last two integrals in Eq. (\ref{test}) require $m_k$ to be fixed; moreover, compatibility between the two 
leads to a condition on quantum numbers of the states that lead to a non-zero expectation value of the 
interaction potential:
\be
m_1 + m_2 = m_3 + m_4.
\label{msum}
\ee
Inspection of the states (\ref{e1} - \ref{e6}) and those beyond shows that many 
states (e.g. \ref{e3} and \ref{e5}) do not contribute 
to the ground state, and hence can be discarded from further discussion.

Furthermore, Gaunt was able to evaluate the $\theta_1$ and $\theta_2$ integrals analytically.\cite{gaunt29} We have 
actually found it simpler to evaluate the integrals as they are (either numerically or analytically with the aid of Maple or
Mathematica); nonetheless Gaunt's formula leads to\cite{slater60} the so-called triangular condition for the angular 
momenta. In the first ($\theta_1$) integration in Eq. (\ref{test}), for example, this requires
\be
\ell_1 + \ell_3 \geq k \geq |\ell_1 - \ell_3|.
\label{k_res}
\ee
Thus, the $k$-sum in Eq. (\ref{test}) is truncated at $k_{\rm max}$, where
\be
k_{\rm max} = min(\ell_1 + \ell_3, \ell_2 + \ell_4).
\ee
It is conventional\cite{slater60,condon35} to introduce coefficients defined as follows:
\begin{widetext}
\begin{eqnarray}
c^k(\ell m;\ell^\prime m^\prime) &=& (-1)^{(m+|m|+m^\prime+|m^\prime| + m - m^\prime + |m-m^\prime|)/2}
\nonumber \\
& & \times \sqrt{(k-|m-m^\prime|)! \over k+|m-m^\prime|)!} \sqrt{(2\ell+1)(\ell-|m|)! \over (\ell +|m|)!} \sqrt{(2\ell^\prime+1)
(\ell^\prime-|m^\prime|)! \over (\ell^\prime+|m^\prime|)!}
\nonumber \\
& & \times {1 \over 2} \int_{-1}^{+1} P_\ell^{|m|}(\mu) P_{\ell^\prime}^{|m^\prime|}(\mu)  P_{k}^{|m - m^\prime|}(\mu) d\mu.
\label{ck}
\end{eqnarray}
In terms of these coefficients the interaction matrix element can be written as
\be
\langle \phi_{i_1} \phi_{i_2} | H_{int} | \phi_{i_3} \phi_{i_4} \rangle = {e^2 \over 4 \pi \epsilon_0} \delta_{m_1+m_2,m_3+m_4} \sum_{k=0}^Q c^k(\ell_1m_1;\ell_3m_3) c^k(\ell_4m_4;\ell_2m_2)R^k(n_1\ell_1;n_2\ell_2;n_3\ell_3;n_4\ell_4),
\label{int_ele}
\ee
where
\be
R^k(n_1\ell_1;n_2\ell_2;n_3\ell_3;n_4\ell_4) =  \int_0^\infty \int_0^\infty R_{n_1,\ell_1}(r_1) R_{n_2,\ell_2}(r_2) R_{n_3,\ell_3}(r_1) R_{n_4,\ell_4} (r_2) {r^k_< \over r^{k+1}_>} r_1^2 r_2^2 dr_1 dr_2
\label{double_r}
\ee
\end{widetext}
This last double integration can be readily done by hand (though it is tedious) or can be done (symbolically) with 
Mathematica or Maple, since the radial wave functions can be readily expressed in terms of associated 
Laguerre polynomials.\cite{griffiths05} Also note the reversed order of the arguments in the second $c^k$ in 
Eq. (\ref{int_ele}); this is important since
\be
c^k(\ell m;\ell^\prime m^\prime) = (-1)^{(m^\prime -m)} c^k(\ell^\prime m^\prime;\ell m).
\label{ckmm}
\ee

Inspection of the states (\ref{e1}-\ref{e6}) and those beyond not already omitted by the condition (\ref{msum}) indicates that three 
distinct possibilities remain (e.g. when $(n_1 \ell_1 m_1) = (n_2 \ell_2 m_2)$ {\em and} 
$(n_3 \ell_3 m_3) = (n_4 \ell_4 m_4)$, etc.) but all of these can be handled through Eq. (\ref{int_ele}).

One other selection rule is present in these results, though not readily apparent. In Eq. (\ref{int_ele}) the {\em 
same value of the index} $k$ must work for both $c^k$ coefficients. These coefficients have been tabulated, for 
example, in the texts by Slater,\cite{slater60} where it is clear that two even or two odd angular momenta (referring 
to $\ell$ and $\ell^\prime$) couple to one another only through even values of $k$, while an even and an odd 
angular momentum couple to one another only through an odd value of $k$. For example $\ell = 0$ and 
$\ell^\prime = 0$  ($ss$) results in a non-zero $c^k$ coefficient only if $k = 0$, while $\ell = 0$ and $\ell^\prime = 2$ 
($sd$) yields a non-zero $c^k$ coefficient only if $k = 2$; the case $\ell = 1$ and $\ell^\prime = 1$ ($pp$) has a 
non-zero $c^k$ coefficient if $k = 0$ or $k = 2$. In contrast the $sp$, $sf$, and $pd$ coefficients are non-zero
for $k=1$, $k=3$, and $k=1,3$, respectively. This means that $\ell_1 + \ell_3$ has to have the same parity as 
$\ell_2 + \ell_4$. Since the parity of the $1s\uparrow 1s\downarrow$ state is even, then, for example, we can further discard 
state (\ref{e4}) from the list, leaving only 3 of the original 6 states listed.

\subsection{An example}

By way of example we quote some steps for the first state (not listed) that does not utilize single particle 
$s$~states, 
\be
\psi_{14}(\vec{r_1},\vec{r_2}) = \frac{1}{\sqrt{2}} \bigl[\phi_{21-1}(\vec{r_1}) \phi_{211}(\vec{r_2}) +
\phi_{21-1}(\vec{r_2}) \phi_{211}(\vec{r_1}) \bigr].
\label{e14}
\ee
To compute the matrix element that couples the first (and primary) basis state with 
this one, we need to evaluate the overlap integral,
\be
H_{1,14} = \langle \psi_1 | H_{\rm int} | \psi_{14}\rangle.
\label{1_14}
\ee
This consists of two overlaps (Eq. (\ref{e14}) has two terms), but, through a change of variables 
$\vec{r_1} \leftrightarrow \vec{r_2}$, these are readily seen to equal one another. We are thus left with
\be
H_{1,14} = \sqrt{2} \langle \phi_{100} \phi_{100} | H_{\rm int} | \phi_{21 -1} \phi_{211} \rangle.
\label{1_14b}
\ee
Eq. (\ref{k_res}) tells us that $1 \geq k \geq 1$, i.e. only $k=1$ need be considered. Then a straightforward evaluation 
of $c^1(00;1 -1)$ and $c^1(11;00)$ (as required in Eq. (\ref{int_ele})) gives us $-1/\sqrt{3}$ for the first and 
$+1/\sqrt{3}$ for the second (using Eq. (\ref{ckmm})). Next only one double radial integral (Eq. (\ref{double_r}) for $k=1$) 
is required, and a straightforward evaluation, readily done by hand, gives
\be
R^1(10;10;21;21) = {112 \over 2187}{Z \over a_0}.
\label{r1}
\ee
Combining this with the $c^1$'s and the $\sqrt{2}$ from Eq. (\ref{1_14b}) gives
\be
H_{1,14} = - \sqrt{2} {448 \over 6561} E_0,
\label{1_14c}
\ee
where $E_0 \equiv {\hbar^2 \over 2 ma_0^2} \approx 13.606$ eV is adopted 
as our unit of energy, and we have used $a_0 = {4\pi \epsilon_0 \over e^2}
{\hbar^2 \over m}$.

In practice we have written a program in Maple to perform these tasks, for the 
bound basis states, up to some cutoff. That is, all the matrix elements, $H_{ij}$, 
up to some cutoff, $i_{\rm max}$, are evaluated, and then this matrix is 
diagonalized. The cutoff $i_{\rm max}$ is defined as the `$n$' quantum number 
up to which all states are included. For example, if $i_{\rm max} = 2$, then 
only 5 basis states are considered. 
The rest of the states either do not contribute 
to the ground state, or, if one of the Hydrogenic single particle states has $n=3$, 
only contributes to the next shell ($i_{\rm max} = 3$) and beyond. Matrix elements
for the case $i_{\rm max} = 2$ are tabulated in the next section.

\subsection{Matrix equation for $I_{\rm max} = 2$}

If we restrict basis states in Section II to those with $n=2$ or less, only 5 two particle
states need to be considered; referring to Eqs. (\ref{e1}-\ref{e6}), these are $\psi_1$,
$\psi_2$, $\psi_6$, $\psi_{14}$, and $\psi_{16}$,
where $\psi_{14}$ is given in Eq. (\ref{e14}), and
\be
\psi_{16}(\vec{r_1},\vec{r_2}) = \phi_{210}(\vec{r_1}) \phi_{210}(\vec{r_2}).
\label{e16}
\ee
Writing the wave function in terms
of these wave functions alone gives rise to a $5 \times 5$ matrix diagonalization problem
(see Eq. (\ref{eig_eqn})). The resulting equation is:
\begin{widetext}
\begin{equation}
\begin{bmatrix} {11 \over 2} & -{32768 \over 64827} & -{64 \over 729} & - {448 \over 6561}\sqrt{2} & -{448 \over 6561}  \\ 
-{32768 \over 64827} & {2969 \over 729} & -{4096 \over 84375} & {2048 \over 28125} \sqrt{2} & {2048 \over 28125} \\
-{64 \over 729} &  -{4096 \over 84375} & {179 \over 128} & -{15 \over 128} \sqrt{2} & -{15 \over 128} \\
- {448 \over 6561}\sqrt{2} & {2048 \over 28125} \sqrt{2} & -{15 \over 128}\sqrt{2} & {47 \over 40} & -{27 \over 640}\sqrt{2} \\
 -{448 \over 6561} & {2048 \over 28125} &  -{15 \over 128} & -{27 \over 640}\sqrt{2}  & {779 \over 640} \end{bmatrix} 
\left[ \begin{array}{c} a_1 \\
                                     a_2 \\
                                     a_{7} \\
                                     a_{14} \\
                                     a_{16} \\  \end{array} \right]  = 
{E \over E_0} 
\left[ \begin{array}{c} a_1 \\
                                     a_2 \\
                                     a_{7} \\
                                     a_{14} \\
                                     a_{16} \\  \end{array} \right] 
\label{5x5}
\end{equation}
\end{widetext}
where $E_0 \approx 13.606$ and the subscripts on the coefficients correspond to the labels in the wave functions.
The resulting ground state energy is $E_1 = -77.13$ eV, and the
eigenvector has components
\begin{eqnarray}
a_{100,100} &=&  \phantom{+}0.9520 \nonumber \\
a_{100,200} &=& -0.3040 \nonumber \\
a_{200,200} &=& -0.0146 \nonumber \\
a_{21-1,211} &=& \phantom{+}0.0267 \nonumber \\
a_{210,210} &=& -0.0188.
\label{imax2}
\end{eqnarray}
Clearly the $\psi_{(100,100)}$ basis state dominates, and small adjustments occur
as the number of basis states increases. Nonetheless, almost 10\% of the wave function
is comprised of components beyond the $1s\uparrow 1s\downarrow$ state.
The matrix construction and diagonalization indicated here is repeated, with the aid of the software package
MAPLE, for increasing values of $I_{\rm max}$, and the results from these calculations are reported in the text.

\section{Hylleraas wave function}

\subsection{The expectation value of the energy}

For reference, the expectation value of the energy, using the normalized Hylleraas wave function 
${1 \over \sqrt{N}}|\Psi_{\rm Hy} \rangle$ with $|\Psi_{\rm Hy} \rangle$ given by Eq. (\ref{hy}), is
\be 
E_{\rm Hy} = - {\hbar^2 \over 2ma_0^2} {4M \over N}(kZ)^2,
\label{e_hy}
\ee
where $k$ can be determined analytically in terms of $c_1$ and $c_2$ [see Eqs. (\ref{params}), below],
\be
kZ = {ZL - L^\prime \over 2M},
\label{k}
\ee
and the unit of energy is the Rydberg,
\be
{\hbar^2 \over 2ma_0^2} \equiv 1\ {\rm Ryd}  \approx 13.606 \ {\rm eV}.
\label{ryd}
\ee
The parameters $L$, $L^\prime$, $M$, and $N$, corresponding to different parts of the Hamiltonian, are given by
\begin{eqnarray}
L = &&4 + 30c_1 + 48c_2 + 72c_1^2 + 280c_1c_2 + 576c_2^2 \nonumber \\
L'=&& 5/4 + 8c_1 + 9c_2 + (35/2)c_1^2 + 48c_1c_2 + 78c_2^2 \nonumber \\
M= &&2 + (25/2)c_1 + 24c_2 + 32c_1^2 + 146c_1c_2 + 480c_2^2  \nonumber \\
N= &&4 + 35c_1 + 48c_2 + 96c_1^2 + 308c_1c_2 + 576c_2^2. 
\label{params}
\end{eqnarray}
Aside from units (most treatments use so-called chemistry units) these all 
agree with results in the literature.\cite{hylleraas29b,schmid87}
Use of the optimally determined parameters, $c_1\approx 0.0803$ and $c_2 \approx 0.0099$, and $k \approx 0.908$
determines the energy via Eq. (\ref{e_hy}).

\subsection{Overlap integrals: bound states}

With a very accurate wave function in hand, we can simply utilize Eq. (\ref{a_n}) to determine the probability of each
basis state in the ground state wave function. The general Hydrogenic bound state can be written as
\beq
\phi_{n,l,m}(r,\theta,\phi)=G_{n,\ell}e^{-Zr/na_0}r^\ell L^{2\ell+1}_{n-\ell-1}\bigg(\frac{2Zr}{na_0}\bigg)Y^m_\ell(\theta,\phi),
\eeq
where
\beq
G_{n,l}=\sqrt{\bigg(\frac{2Z}{na_0}\bigg)^3\frac{(n-\ell-1)!}{2n[(n+\ell)!]^3}}\bigg(\frac{2Z}{na_0}\bigg)^\ell, 
\eeq
and the required overlap integrals consist only of terms of the form
\beq
I=\int\phi_{n_1\ell_1m_1}^*\phi_{n_2\ell_2m_2}^*{1 \over \sqrt{N}}\Psi_{\rm Hy}.
\label{overlapb}
\eeq
In the $u,s,t$ coordinates defined above, and using the volume element,
\beq
{\rm d}\tau=\frac{a_0^6}{8}(s^2-t^2)u \ {\rm d}s{\rm d}t{\rm d}u\ \sin\theta_1{\rm d}\theta_1{\rm d}\phi_1{\rm d}\phi_2,
\eeq
many of these integrals are straightforward to calculate analytically --- an example 
will be shown below. However, it is of interest to compute these overlaps
for general quantum numbers, in particular to check on selection rules, and 
to examine the convergent behaviour for large $n_1$ and $n_2$. To this end
we use the same expansion, Eq. (\ref{multipole}) used earlier to evaluate 
matrix elements of the Hamiltonian, and then evaluate Eq. (\ref{overlapb}). The
details of the derivation for a general matrix element are a little 
too cumbersome to include here; all of the 
overlaps involving bound state wave functions can be expressed in terms of 
Hypergeometric functions, which are readily evaluated using Mathematica, and 
with these we can readily sum the contributions from the bound states.

For students, however, it is more instructive to compute this overlap `by hand' for 
a few of the most relevant states. We outline the procedure for the most important state, 
$\psi_1(\vec{r_1},\vec{r_2}) = \phi_{100}(\vec{r_1}) \phi_{100}(\vec{r_2})$.
Then
\begin{equation}
a_1=\int d^3{r_1} \int d^3r_2\ \phi_{100}(\vec{r_1})\phi_{100}(\vec{r_2}){1 \over \sqrt{N}}\Psi_{\rm Hy}(\vec{r_1},\vec{r_2}),
\label{overlap1}
\end{equation}
and the only angular dependence occurs in the $|\vec{r_1} - \vec{r_2}|$ term in the Hylleraas wave function.
This requires knowledge of the angle between the two vectors, $\vec{r_1}$ and $\vec{r_2}$, which we can call
$\theta_{12}$; this is given as\cite{hylleraas29b}
\begin{equation}
\cos{\theta_{12}} = \cos{\theta_1} \cos{\theta_2} + \sin{\theta_1} \sin{\theta_2} \cos{(\phi_2-\phi_1}),
\label{coslaw}
\end{equation}
where $\vec{r_i} \equiv (r_i,\theta_i,\phi_i)$, for $i=1,2$ are the spherical coordinates for each vector.
A simpler trick is to line up the $z_2$ axis with $\vec{r_1}$; then $\theta_{12} \equiv \theta_2$, 
and all six integrations are then straightforward. The result is
\begin{eqnarray}
a_1&=&{32 \over \sqrt{N}} {k^3 \over (k+1)^6}\biggl\{ 4 + 35c_1 {k \over k+1} 
+ 96 c_2 \bigl({k \over k+1}\bigr)^2 \biggr\}, \nonumber \\
&\approx& 0.9624,
\label{overlap2}
\end{eqnarray}
and therefore $a_1^2 \approx 0.9262$.
The numerical values follow upon substitution of the optimal values of $k$, $c_1$, and $c_2$.
Similar calculations can be performed for the other overlap coefficients; some results are tabulated in Table 1,
where the constituents of the basis states, suitably symmetrized, are listed. Note that the label provided under the
`basis state' column identifies the two single particle wave functions involved, and the 
basis state is a singlet state, and therefore the spatial part is symmetrized, as 
enumerated in Eqs. (\ref{e1}-\ref{e6}).

\section{Evaluation of overlap integrals}

We provide some of the details for the evaluation of Eq. (\ref{expansion}). Because
of the symmetry in the singlet state , and using the definition
\begin{equation}
a_p = \sqrt{2 \over N} {2 \over \pi} \bigl( {kZ \over a_0} \bigr)^3 A_p
\label{aa1}
\end{equation}
we require the integral
\begin{widetext}
\begin{equation}
A_p = \int d^3r_1 \int d^3r_2 \phi_p(r_1) \phi_{100}(r_2) e^{-{Z k \over a_0}(r_1+r_2)}
\bigl\{
1 + 2 Zc_1 {k \over a_0}|\vec{r_1} - \vec{r_2} | + c_2\bigl[  2Z{k \over a_0}(r_1 - r_2)  \bigr]^2
\bigr\},
\label{aa2}
\end{equation}
where
\begin{equation}
\phi_p(r_1) = {1 \over \sqrt{4 \pi}}{Z \over a_0} \sqrt{2 \pi {pa_0 \over Z} \over 1 -
 e^{-2 \pi Z/(pa_0)}}e^{-ipr}M(1 + i{Z \over pa_0},2,2ipr),
\label{aa3}
\end{equation}
and M(a,b,z) is the Kummer function.\cite{abramowitz72}
Note that
\begin{equation}
\int_0^{\pi} d\theta_1 \sin{\theta_1} \int_0^{\pi} d\theta_2 \sin{\theta_2} 
\int_0^{2\pi} d\phi_1 \int_0^{2\pi} d\phi_2 \ [1] = (4\pi)^2,
\label{aa4}
\end{equation}
while
\begin{equation}
\int_0^{\pi} d\theta_1 \sin{\theta_1} \int_0^{\pi} d\theta_2 \sin{\theta_2} 
\int_0^{2\pi} d\phi_1 \int_0^{2\pi} d\phi_2 |\vec{r_1} - \vec{r_2}| = 
(4\pi)^2\bigl(r_> + {1 \over 3}{r_<^2 \over r_>}
\bigr),
\label{aa5}
\end{equation}
\end{widetext}
where $r_<$ ($r_>$) refers the smaller (larger) of $r_1$ or $r_2$. The first of these
is trivial, while the second follows most readily by using the trick mentioned following
Eq. (\ref{coslaw}).

Two integrations remain, and it is efficient to switch to dimensionless variables, $x_1 \equiv
Zr_1/a_0$, $x_2 \equiv Zr_2/a_0$, and $y \equiv pa_0/Z$. Then, since the Kummer function
depends only on $x_1$, the $x_2$ integration is elementary, and leaves behind polynomials
in $x_1$. One arrives at the expression given by Eq. (\ref{expansion2}), with the definitions
\begin{eqnarray}
I_n(y) &\equiv & B_n(y,k+iy) \phantom{aaaa} {\rm and} \nonumber \\
J_n(y) &\equiv & B_n(y,2k+1+iy),
\label{iandj}
\end{eqnarray}
where
\begin{equation}
B_n(y,z) \equiv \int_0^\infty dx \ x^n e^{-zx}M(1+{i \over y},2,2iyx).
\label{bdefn}
\end{equation}
Now an expansion of the Kummer function (following Eq. (\ref{kummer})) allows
us to do the integral and obtain an infinite summation which can be recognized as
a Hypergeometric function,\cite{abramowitz72}
\begin{equation}
B_n(y,z) ={n! \over z^{n+1}} F[1+{i \over y},n+1;2;{2iy \over z}],
\label{beval}
\end{equation}
with $n=1,2,3,4$ needed. The standard definition of the Hypergeometric function
uses the Gauss hypergeometric series,\cite{abramowitz72} 
with circle of convergence in the unit circle $|z|=1$
\begin{equation}
F(a,b;c;z) = \sum_{m=0}^\infty {(a)_m (b)_m \over (c)_m} {z^m \over m!},
\label{hyper}
\end{equation}
where the $(a)_m$ are the Pochhammer symbols introduced earlier.
Writing Eq. (\ref{beval}) wouldn't normally be too useful (because Hypergeometric
functions are hard to evaluate numerically), except that one can use the identity\cite{abramowitz72}
\begin{equation}
F[a,b;c;z] \equiv (1-z)^{c-a-b} F[c-a,c-b;c;z].
\label{hyper_identity}
\end{equation}
This identity is extremely helpful because the second variable in the Hypergeometric function
on the right hand side is a non-positive integer for
$n=1,2,3,4$, and so, because of the definition of the Pochhammer symbol, the infinite
sum in the definition of the Hypergeometric function becomes truncated to 
$n$ terms.
We thus obtain, for $n=1,2,3,4$,
\begin{equation}
I_n(y) = {1 \over k^2 + y^2} \exp{\bigl(-{2 \over y}{\rm tan}^{-1}{y \over k}\bigr)} D_n
\label{ineval}
\end{equation}
and
\begin{equation}
J_n(y) = {1 \over (2k+1)^2 + y^2} \exp{\bigl(-{2 \over y}{\rm tan}^{-1}({y \over 2k+1})\bigr)} E_n,
\label{jneval}
\end{equation}
where
\begin{equation}
D_n = {n! \over (k-iy)^{n-1}} F[1-{i \over y},1-n;2;{2iy \over k+iy}],
\label{dneval}
\end{equation}
and
\begin{equation}
E_n = {n! \over (2k+1-iy)^{n-1}} F[1-{i \over y},1-n;2;{2iy \over 2k+1+iy}].
\label{eneval}
\end{equation}
With these definitions, straightforward evaluation gives
\begin{eqnarray}
D_1 &=& 1 \nonumber \\
D_2 &=& {2(k-1) \over k^2 + y^2} \nonumber \\
D_3 &=& {4(k-1)(2k-1) \over (k^2+y^2)^2} - {2 \over k^2+y^2} \nonumber \\
D_4 &=&{8(k-1)(2k-1)(3k-1) \over (k^2+y^2)^3} - {8(3k-2) \over (k^2+y^2)^2}
\label{d_n}
\end{eqnarray}
and
\begin{eqnarray}
E_1 &=& 1 \nonumber \\
E_2 &=& {4k \over (2k+1)^2 + y^2}.
\label{e_n}
\end{eqnarray}
Summarizing, we have
\begin{eqnarray}
I_1 &=& {1 \over k^2 + y^2} \exp{\bigl(-{2 \over y}{\rm tan}^{-1}{y \over k}\bigr)} \nonumber \\
I_2 &=& {2(k-1) \over (k^2 + y^2)^2} \exp{\bigl(-{2 \over y}{\rm tan}^{-1}{y \over k}\bigr)} \nonumber \\
I_3 &=& \biggl\{ {4(k-1)(2k-1) \over (k^2+y^2)^3} - {2 \over (k^2+y^2)^2} \biggr\} \exp{\bigl(-{2 \over y}{\rm tan}^{-1}{y \over k}\bigr)} \nonumber \\
I_4 &=& \biggl\{ {8(k-1)(2k-1)(3k-1) \over (k^2+y^2)^4} - {8(3k-2) \over (k^2+y^2)^3} \biggr\} \exp{\bigl(-{2 \over y}{\rm tan}^{-1}{y \over k}\bigr)} \nonumber \\
J_1 &=& {1 \over (2k+1)^2 + y^2} \exp{\bigl(-{2 \over y}{\rm tan}^{-1}({y \over 2k+1})\bigr)} \nonumber \\
J_2 &=& {4k \over ((2k+1)^2 + y^2)^2} \exp{\bigl(-{2 \over y}{\rm tan}^{-1}({y \over 2k+1})\bigr)}
\label{iandj2}
\end{eqnarray}
and these are to be substituted into Eq.(\ref{expansion2}).


\begin{thebibliography}{1}

\bibitem{hylleraas29a}E.A. Hylleraas, ``On the Ground State of the Helium Atom,'' Z. Phys. {\bf 48}, 469 (1929). The work in this
paper by Dr. Biem\"uller was never published, to our knowledge. Hylleraas' paper and many others are reprinted in
English in Ref. [\onlinecite{hettema00}].

\bibitem{hettema00} H. Hettema, {\it Quantum Chemistry: Classic Scientific Papers} (World Scientific, Singapore, 2000).

\bibitem{hylleraas29b}E.A. Hylleraas,``A New Calculation of the Energy of Helium in the Ground State as Well 
as the Lowest Term of Ortho-helium, Z. Phys. {\bf 54}, 347 (1929). This is also reprinted in English in 
Ref. [\onlinecite{hettema00}]. See also {\it Mathematical and Theoretical Physics, Volume II}, by 
E.A. Hylleraas (Wiley-Interscience, Toronto, 1970), particularly Chapter 7.

\bibitem{rem1} We use the phrase, ``singlet state $1s\uparrow 1s\downarrow$'' to refer to the spin wave function
${1 \over \sqrt{2}}\bigl[1s\uparrow 1s\downarrow - 1s\downarrow 1s\uparrow \bigr]$.

\bibitem{rem2} That a $1s\uparrow 1s\downarrow$-like state  (i.e. a non-Hydrogenic $S$ state with zero total angular 
momentum) will ultimately succeed was recognized early by Eugene Wigner, as recorded in 
E.A. Hylleraas, ``Reminisences of Early Quantum Mechanics  of Two-Electron Atoms,'' Rev. Mod. Phys {\bf 35}, 421 (1963).

\bibitem{schwartz06} See, for example, C. Schwartz, ``Experiment and Theory in Computations of the He Atom Ground State,'' 
International Journal of Modern Physics E {\bf 15}, 877 (2006); in the update (arXiv:math-ph/0605018v1),  
the ground state energy is determined with an accuracy of 50 digits.

\bibitem{rem3} We should note that this work was started {\em without} knowledge of Hylleraas' remark concerning Dr.
Biem\"uller quoted in the opening paragraph, and we initially expected that convergence would be attained with
inclusion of a few shells beyond the singlet state $1s\uparrow 1s\downarrow$' configuration.

\bibitem{rem4} Again, for those that have reflected on this problem, the need to include continuum (or scattering) states
comes as no surprise, and they may even be inclined to dismiss the entire exercise as `bad results due to starting with
a poor basis set.' But we re-iterate, the `poor basis set' with which we start is the one with which we tend to think, and therefore
these questions are worth answering. Furthermore, as outlined in the subsequent discussion, in some research-type 
problems we are forced to adopt a poor basis set, and then the present analysis may provide some useful guidance
for these problems.

\bibitem{hirsch01} J. E. Hirsch, ``Dynamic Hubbard Model,'' Phys. Rev. Lett. {\bf 87}, 206402-1-4 (2001); ``Quantum 
Monte Carlo and exact diagonalization study of a dynamic Hubbard model,'' Phys. Rev. B {\bf 65}, 214510-1-16 (2002);
``Quasiparticle undressing in a dynamic Hubbard model: Exact diagonalization study,'' Phys. 
Rev. B {\bf 66}, 064507-1-12 (2002).

\bibitem{bach10} G. H. Bach, J. E. Hirsch and F. Marsiglio, ``Two-site dynamical mean field theory for the dynamic 
Hubbard model,'' Phys. Rev. B {\bf 82}, 155122-1-15 (2010). 

\bibitem{bach12} G. H. Bach and F. Marsiglio, ``Optical conductivity for a dimer in the dynamic Hubbard model,'' Phys. 
Rev. B {\bf 85}, 155134-1-10 (2012). 

\bibitem{hubbard63} J. Hubbard, ``Electron Correlations in Narrow Energy Bands,'' Proc. R. Soc. London {\bf A 276}, 
238-257 (1963); 
``Electron Correlations in Narrow Energy Bands. II. The Degenerate Band Case,'' {\bf 277}, 237-259 (1964); 
``Electron Correlations in Narrow Energy Bands. III. An Improved Solution.'' {\bf 281}, 401-419 (1964); 
``Electron Correlations in Narrow Energy Bands. IV. The Atomic Representation,'' {\bf 285}, 542-560 (1965).

\bibitem{rem5} A search through APS journals alone reveals over 5000 articles that contain the keyword `Hubbard' in 
their title or abstract.

\bibitem{rem6} This is reminiscent of the situation in General Relativity, where the presence of mass alters the space-time
structure in which the mass is situated.

\bibitem{rem7} Actually the method used to determine the `exact' ground state energy and wave function 
(e.g. in Ref. [\onlinecite{schwartz06}]) is a straightforward generalization of the simplest Hylleraas wave 
function quoted here.

\bibitem{levine09} Ira N. Levine, {\em Quantum Chemistry}, 6th Edition (Pearson, Upper Saddle River, 2009).

\bibitem{griffiths05} D. J. Griffiths, {\it Introduction to Quantum Mechanics} (Pearson/Prentice Hall, Upper Saddle River, NJ, 2005), 2nd edition.

\bibitem{rem8} Ironically, we use particle labels to avoid particle labelling, i.e. to enforce indistinguishability. See 
Footnote \cite{rem1} for equivalent expressions of the spin singlet state.

\bibitem{marsiglio09} F. Marsiglio, ``The harmonic oscillator in quantum mechanics: a third way,'' Am. J. Phys. {\bf 77}, 253 (2009).

\bibitem{schmid87} E.W. Schmid, G. Spitz, and W. L\"osch, {\it Theoretical Physics on the Personal Computer}, (Springer Verlag, New York, 1987), Chap. 16.

\bibitem{abramowitz72} M. Abramowitz and I.A. Stegun, {\it Handbook of Mathematical Functions with
Formulas, Graphs, and Mathematical Tables} (Dover Publications, Inc. New York, 1972).

\bibitem{landau65} L.D. Landau and E.M. Lifshitz, {\em Quantum Mechanics}, 2nd Edition, (Pergamon Press, New York, 1965), Eq. (33.4).

\bibitem{jackson75} J.D. Jackson, {\it Classical Electrodynamics}, 2nd Edition (John Wiley and Sons, Toronto, 1975).

\bibitem{slater60} J.C. Slater, {\it Quantum Theory of Atomic Structure}, (McGraw-Hill, Toronto, 1960). The required formula is found on p. 308.

\bibitem{gaunt29} J.A. Gaunt, ``The Triplets of Helium,'' Phil. Trans. Roy. Soc. (London) {\bf A228}:151 (1929).

\bibitem{condon35} E.U. Condon and H.H. Shortley, {\it Theory of Atomic Spectra}, (Cambridge University Press, New York, 1935).

\end{thebibliography}
\end{document}